\documentclass[journal=jctcce,manuscript=article]{achemso}
\usepackage[version=4]{mhchem}
\usepackage{graphicx}
\usepackage{subcaption}
\usepackage{booktabs}
\usepackage{cleveref}
\usepackage{xcolor}
\usepackage[utf8]{inputenc}
\usepackage{pgfplots}

\usepackage{graphicx} \usepackage{subcaption} 

\usepackage{subcaption} 

\usepackage{natbib}
\usepackage{mathtools}

\usepackage{listings}
\usepackage{xcolor} 

\lstdefinestyle{pyclean}{
    language        = Python,
    basicstyle      = \ttfamily\small,      
    numbers         = left,                 
    numberstyle     = \tiny\color{gray},
    stepnumber      = 1,
    numbersep       = 5pt,
    frame           = single,               
    backgroundcolor = \color{white},
    keywordstyle    = \color{blue},
    commentstyle    = \color{gray!70}\itshape,
    stringstyle     = \color{orange},
    showstringspaces= false,
    tabsize         = 4,
    columns         = flexible,
    keepspaces      = true,                 
    breaklines      = false,                
    linewidth       = \linewidth,
    xleftmargin     = 1.5em,                
    xrightmargin    = 1.5em,
    aboveskip       = 0.5\baselineskip,
    belowskip       = 0.5\baselineskip,
}

\title{User-defined Electrostatic Potentials in DFT Supercell Calculations: Implementation and Application to Electrified Interfaces	}

\author{Samuel Mattoso}
\affiliation{ 
Max Planck Institute for Sustainable Materials, 
Max-Planck-Straße 1, 40237 Düsseldorf, Düsseldorf, Germany}

\author{Jing Yang}
\affiliation{ 
Max Planck Institute for Sustainable Materials, 
Max-Planck-Straße 1, 40237 Düsseldorf, Düsseldorf, Germany}

\author{Florian Deißenbeck}
\affiliation{ 
Max Planck Institute for Sustainable Materials, 
Max-Planck-Straße 1, 40237 Düsseldorf, Düsseldorf, Germany}

\author{Ahmed Abdelkawy}
\affiliation{ 
Max Planck Institute for Sustainable Materials, 
Max-Planck-Straße 1, 40237 Düsseldorf, Düsseldorf, Germany}

\author{Christoph Freysoldt}
\affiliation{ 
Max Planck Institute for Sustainable Materials, 
Max-Planck-Straße 1, 40237 Düsseldorf, Düsseldorf, Germany}

\author{Stefan Wipperman}
\affiliation{ 
Philipps-Universität Marburg,
 Renthof 5, 35032 Marburg, Germany}

\author{Mira Todorova}
\affiliation{ 
Max Planck Institute for Sustainable Materials, 
Max-Planck-Straße 1, 40237 Düsseldorf, Düsseldorf, Germany}
\email{m.todorova@mpi-susmat.de}

\author{Jörg Neugebauer}
\affiliation{ 
Max Planck Institute for Sustainable Materials, 
Max-Planck-Straße 1, 40237 Düsseldorf, Düsseldorf, Germany}

\begin{document}

\begin{abstract}
Introducing electric fields into density functional theory (DFT) calculations is essential for understanding electrochemical processes, interfacial phenomena, and the behavior of materials under applied bias. However, applying user-defined electrostatic potentials in DFT is nontrivial and often requires direct modification to the specific DFT code. In this work, we present an implementation for supercell DFT calculations under arbitrary electric fields and discuss the required corrections to the energies and forces. The implementation is realized through the recently released VASP–Python interface, enabling the application of user-defined fields directly within the standard VASP software and providing great flexibility and control. We demonstrate the application of this approach with diverse case studies, including molecular adsorption on electrified surfaces, field ion microscopy, electrochemical solid–water interfaces, and implicit solvent models. 

\end{abstract}

\section{Introduction}

Simulating the response of molecules and materials to external electrostatic potentials and fields is of fundamental importance to chemistry, physics, biology and materials science \cite{English2015-xr}. External electric fields may originate from various environmental sources, including solvents\cite{dipino2025deconstructingoriginsinterfacialcatalysis,Jang2022-jo}, electrodes \cite{Sundararaman2022-vg,todorova2024principlesapproachesconceptselectrochemical,Gonella2021-to}, charged tips from measuring instruments such as in atom probe tomography \cite{10.1093/micmic/ozad067.291}, from biological membranes and ion channels\cite{Nishihara2017-wf,Roux2004-rv}. The role these fields play is critical, as they can drive chemical reactions\cite{Surendralal2018-ug,Deisenbeck2024-fk,WARSHEL1976227}, induce modifications to surfaces \cite{Katnagallu2025-aq}, influence the folding of proteins \cite{Bellissent-Funel2016-lf} and determine the  selectivity of reaction pathways \cite{Yu2021-md,Yang2023-rv}.

Density Functional Theory (DFT) is commonly used to capture the quantum mechanical behavior of molecules and materials under the effect of an external field, due to its favorable balance between computational cost and accuracy\cite{Butera2024-uv,Mardirossian2017-uj}. DFT calculations can be performed with a variety of electronic structure codes, such as the Vienna Ab Initio Simulation Package (VASP)\cite{Kresse1996-gq,Hafner1997-ix}, ORCA\cite{Neese2012-sk}, QUANTUM ESPRESSO\cite{Giannozzi2009-rk}, Molpro\cite{Werner2012-we}, GPAW \cite{Enkovaara2010-hs}, S/PHI/nX\cite{Boeck2011,Freysoldt2020} and many others. These electronic structure codes support the use of external fields and potentials with varying flexibility, from the use of point charges, external plugin files, in-code commands or relying on direct modification of the base code from the user. Well-established, built-in uses of external potentials include the dipole correction \cite{PhysRevB.46.16067} or implicit solvent methods \cite{Ringe2021,Mathew2014}. Beyond these, ongoing efforts seek other external field capabilities, such as for modeling electric fields in electrochemistry or solvation effects when referring to quantum mechanics/molecular mechanics (QM/MM) approaches\cite{Lim2016,Govindarajan2025-hv}.

Implementing these external electrostatic potentials in practice has often required direct modification of the source codes of electronic structure packages by users or research groups. Although such efforts have lead to progress on specific scientific applications, they demand intimate knowledge of large and complex code bases and are additionally prone to the introduction of errors, difficult to reproduce or maintain across software versions. Custom patches hinder portability and subsequent extensions by others, and ad hoc edits to individual subroutines can inadvertently cause bugs and regressions in other parts of the code. These challenges highlight the need for modular, well‑documented interfaces between the source code and the user, that allow for the interior modification of the complex electronic structure codes, while insulating their core functionality.

The freedom to modify selected components and quantities in the source code, gives rise to a high flexibility of computational setups and systems that can be explored. Nevertheless, careful validation of the computational setup is essential to ensure correct electrostatic boundary conditions, in addition to the physical consistency of the forces and energies under the effect of an external potential. In practical terms, this entails ensuring charge neutrality and avoiding simulation artifacts due to long-range Couloumb interactions, thermodynamic inconsistencies and so forth.
Additionally, correction terms must be added to the forces and energies, due to their electrostatic interactions with the external field. Lastly, it is important to note that upon modification of the software package, the responsibility of ensuring numerical and physical correctness shifts from the software developer to the user.

In this context, we discuss the theoretical and computational details for the setup of arbitrary electrostatic potentials in DFT codes while ensuring correct forces, energies and electrostatic boundary conditions. For this purpose, we employ the recently released  VASP 6.5.0  version\cite{Vasp650}, which allows for the straightforward modification of the local potential and other internal quantities within the source code through a Python scripting interface. 
We make use of these features to apply electric fields across interfaces, to control the voltage in electrochemical simulations, and to impose an external solvation potential. These methodologies enable a wide range of applications, such as the modeling of  solid-liquid interfaces, electrocatalysis, solvation effects and materials. Beyond the physical and numerical consistency, the proposed computational setup between the Python plugin and the electronic structure source code ensures robustness, modularity and extensibility.




\section{Theoretical and Computational Details}


\subsection{Work Flow of the VASP-Python Interface}

\begin{figure}
  \centering
  \begin{subfigure}[b]{0.65\textwidth}
    \includegraphics[width=\textwidth]{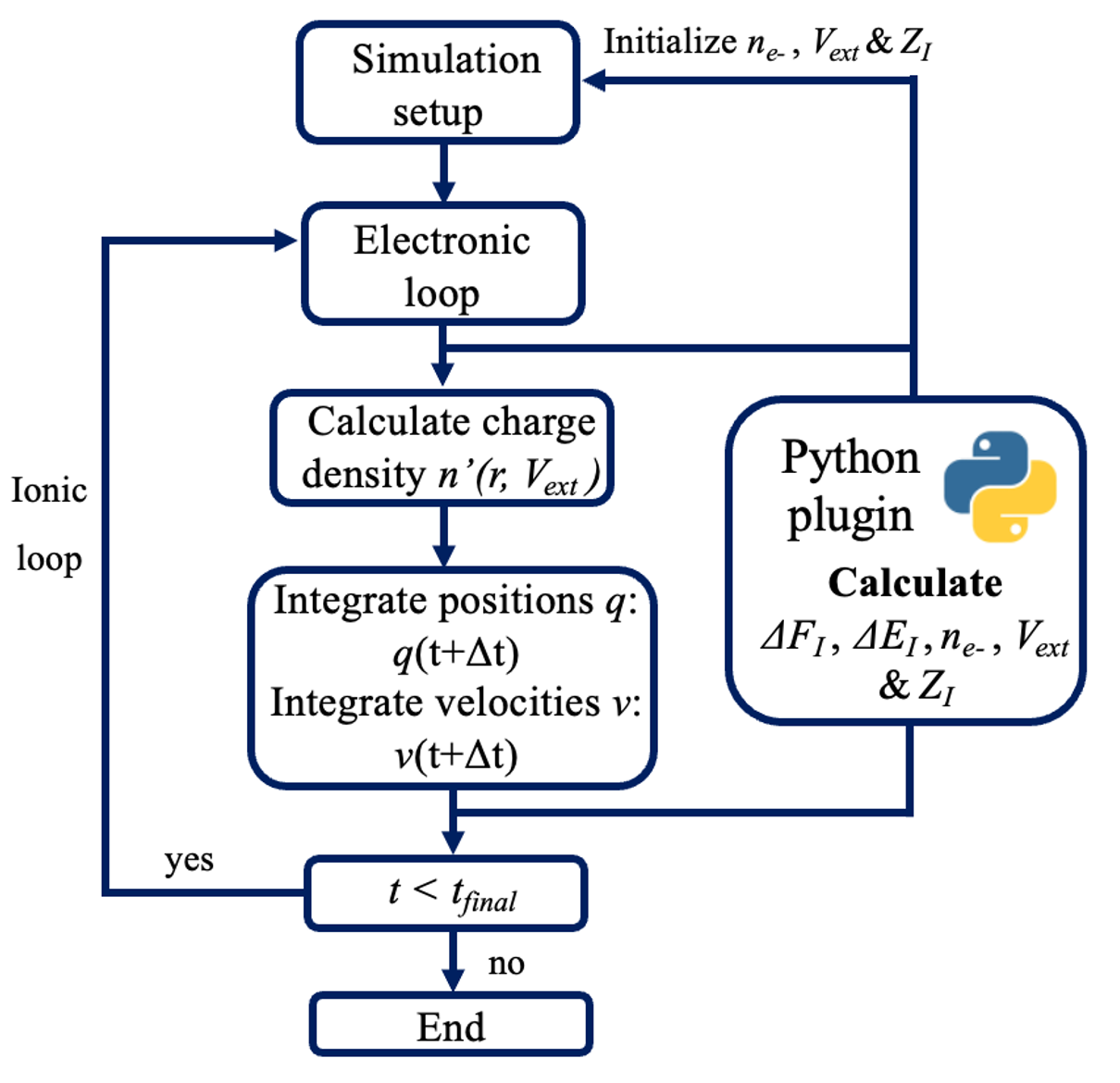}

  \end{subfigure}
  \hfill
    \caption{Flowchart of the VASP-Python plugin setup during a Molecular Dynamics-DFT simulation. $\Delta F_I $, $\Delta E_I $, $n_{e-}$, $V_{ext}$ and $Z_{I}$ refer to the forces, total energies, number of electrons, external potential and nuclear charges, respectively. }
  \label{fig:schematic picture}
\end{figure}

The Python interface introduced in VASP provides direct access to the local potential, nuclear charges and forces, as well as total energies. These quantities can be modified by the Python script during initialization and dynamically during the ionic loop for an energy minimization or molecular dynamics run.
A schematic overview of the workflow between the electronic structure code and the Python plugin is shown in  Figure \ref{fig:schematic picture}. For the purpose of applying an external electrostatic potential, $V_{ext}$ is added to the total potential during the electronic loop. As such, the charge density $n'(\mathbf{r},V_{ext})$ under the applied bias  is calculated by the VASP code. However, by design of the VASP-Python interface, the interaction between $V_{ext}$ and the nuclear cores are not included automatically. In order to obtain the correct energies and forces, these contributions (force correction $\Delta F_I$ and energy correction $\Delta E_I$, where the subscript $I$ represents the $I$th atom) should also be added through the Python plugin, which we discuss in Sec. \ref{sec:energy_force}.   

Additionally, the VASP-Python plugin enables changes to the number of electrons ($n_e$), as well as the nuclear core charge ($Z_I$), for each ionic step. This enables the application of an external electric field that changes with time during an \textit{ab initio}  molecular dynamics (AIMD) run. An important example where this functionality is useful is potential-controlled simulations of electrochemical interfaces, which we discuss in more detail in Sec. \ref{sec:electrochemistry}.

In this work, we refer only to quasi-static fields that can be modeled within the framework of the Born-Oppenheimer approximation. This contrasts with time-dependent oscillating fields, such as those due to electromagnetic waves, which are relevant for simulating spectroscopic phenomena.

\subsection{Energy and Force Corrections}
\label{sec:energy_force}

In this section, we derive the modifications of the total energy and forces under the external electrostatic field and identify the terms that need to be included through the VASP-Python plugin. The derivation is adapted from applying the dipole correction to DFT calculations \cite{PhysRevB.46.16067}. The energy functional without $V_{ext}$ can be written as
\begin{equation}
    E[n(\mathbf{r})] = T[n] + E^{e\text{-}e}[n] + \int V^{ion}(\mathbf{r})n(\mathbf{r})d\mathbf{r} + E^{ion\text{-}ion}\quad,
    \label{eq:energy-functional}
\end{equation}
where $T[n]$, $E^{e\text{-}e}[n]$, and $E^{ion\text{-}ion}$ represent the kinetic, electron-electron, and ion-ion interaction energy. $V^{\mathrm{ion}}$ is the ionic pseudopotential. Upon the application of $V_{ext}$, the energy functional changes to
\begin{equation}
    \tilde{E}[n(\mathbf{r})] = E[n(\mathbf{r})] + \int V_{ext}(\mathbf{r}) n(\mathbf{r}) d\mathbf{r} + \int V_{ext}(\mathbf{r}) n^{ion}(\mathbf{r})d\mathbf{r} \quad.
    \label{eq:hamiltonian}
\end{equation}
Here $n^{ion}(\mathbf{r})$ represents the density of the nuclear core charges. They are represented as point charges in VASP. 

The two terms added in Eq. \ref{eq:hamiltonian} represent the interaction between the electron charge with the external field and the nucleus charge with the external field, respectively. When $V_{ext}$ is added to the total potential in VASP, $n(\mathbf{r})$ relaxes self-consistently to $n'(\mathbf{r},V_{ext})$. As such, VASP also computes the energy functional as $E[n'(\mathbf{r})]$, including the effect of $V_{ext}$ to all terms in Eq. \ref{eq:energy-functional}. However, for the two additional terms in Eq. \ref{eq:hamiltonian}, by default the VASP output energy includes only the interaction between $V_{ext}$ and the electron density but not the nucleus. Therefore an energy correction needs to be added for the $I$th nucleus as 
\begin{equation}
    \Delta E_I = -Z_{I}V_{ext}(\mathbf{R}_I) \quad.
    \label{eq:energy_correction}
\end{equation}

Similarly, in the VASP force output, the interaction between $V_{ext}$ and the nucleus charge is also lacking, and a force correction should be added to each nucleus 
\begin{equation}
\Delta F_I = -\frac{\partial \Delta E_I}{\partial \mathbf{R}_I} 
= -Z_I \mathcal{E}^{\mathrm{ext}}(\mathbf{R}_I) \quad ,
\label{Eq:force_correction}
\end{equation}
where $\mathcal{E}^{\mathrm{ext}}$ represents the externally applied electric field corresponding to $V_{ext}$. We note that the derivations here for incorporating the external potential into the Hamiltonian is general to all DFT codes, not limited to VASP.

\begin{figure}
    \centering
    \begin{subfigure}[b]{0.44\textwidth}
        \caption{}
        \includegraphics[width=\textwidth]{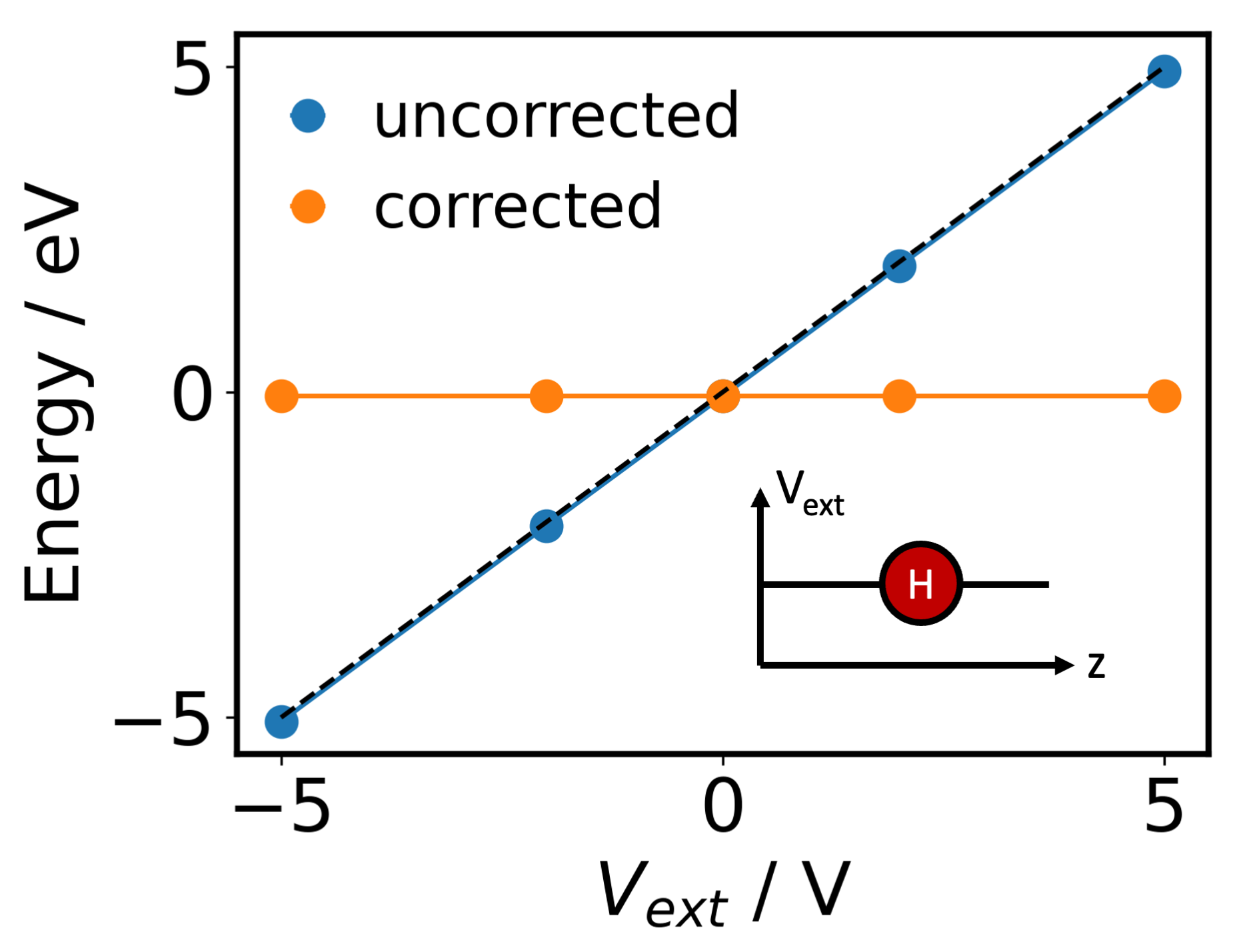}
    \end{subfigure}
    \begin{subfigure}[b]{0.48\textwidth}
        \caption{}
        \includegraphics[width=\textwidth]{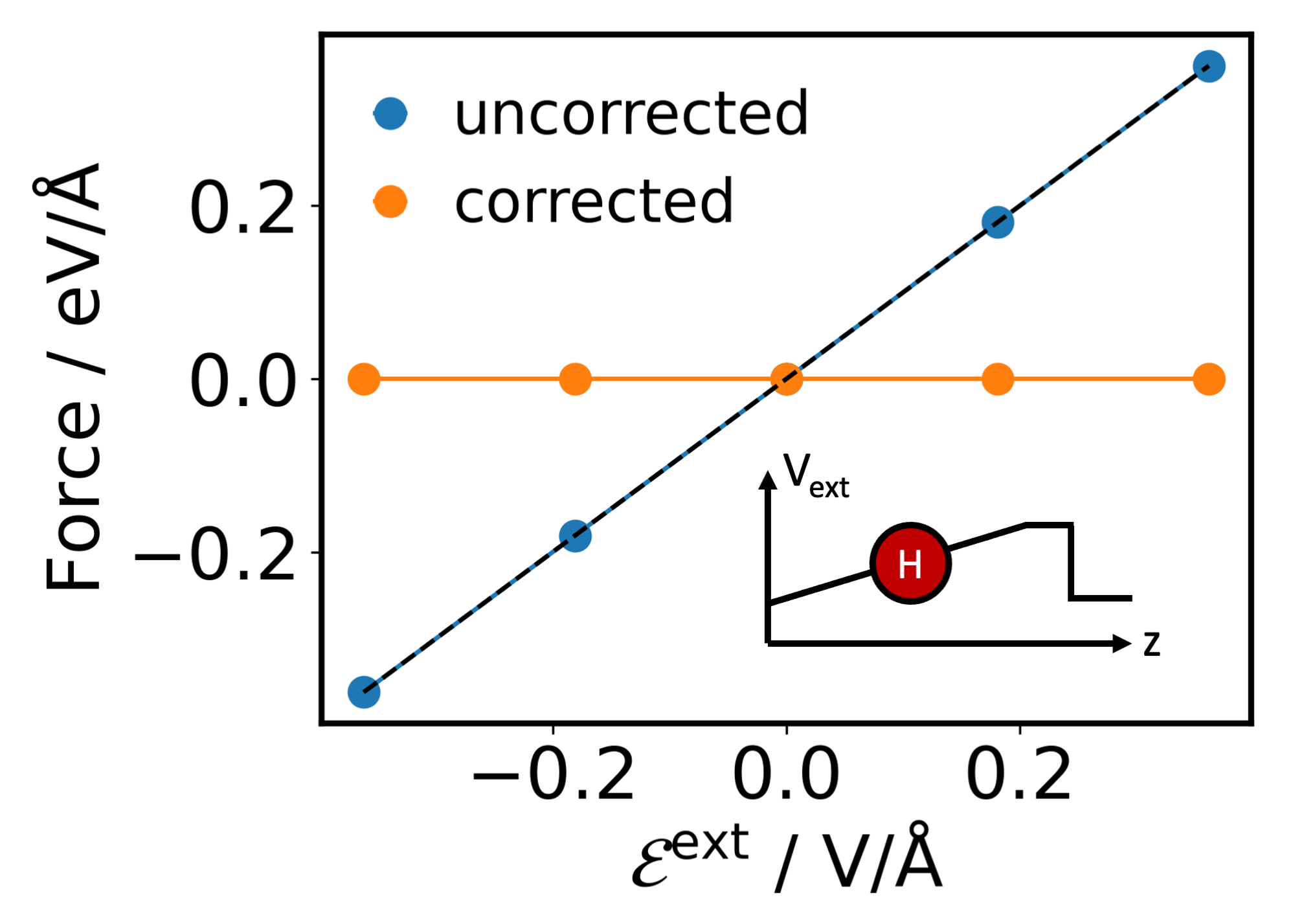}
    \end{subfigure}
    \caption{
        (a) Energy of a hydrogen atom in a constant potential $V_{ext}$ before and after the core correction (see text). The black dashed line represents the energy equal to $1e\cdot V_{ext}$. (b) The force in the $z$ direction of a hydrogen atom in a linear potential corresponding to a constant field $\mathcal{E}^{\mathrm{ext}}$ before and after the correction. The black dashed line represents the force equal to $1e\cdot \mathcal{E}^{\mathrm{ext}}$. }
    \label{fig:H}
\end{figure}

To benchmark the proposed energy and force correction scheme, we perform tests on the model system of a neutral hydrogen atom in vacuum (Fig. \ref{fig:H}). In Fig. \ref{fig:H}a, a constant potential is added through the VASP-Python interface. Since the system is charge neutral, it is expected that the energy of the system does not change. However, the uncorrected energy (blue line) shows a linear dependence on $V^{ext}$, which scales with $1e\cdot V^{ext}$ (black dashed line). This behavior matches Eq. \ref{eq:energy_correction}, as the $Z_I$ value for hydrogen is 1. After the energy correction, the energy of the hydrogen atom is invariant with respect to $V^{ext}$. 

Similarly, in Fig. \ref{fig:H}b, we test the force correction by putting the hydrogen atom in a constant electric field $\mathcal{E}^{\mathrm{ext}}$. As the hydrogen atom is charge-neutral and its dipole moment due to field polarisation is negligible, the force should be close to zero. However, again we observe that the uncorrected force (blue line) has a strong linear dependence on $\mathcal{E}^{\mathrm{ext}}$. The dependence follows the line $1e\cdot \mathcal{E}^{\mathrm{ext}}$ (black dashed line), as derived in Eq. \ref{Eq:force_correction}. These tests demonstrate that the interaction between the nuclear core charges and the external potential is indeed not included in the VASP-calculated energies and forces, and that the corrections in  Eq. \ref{eq:energy_correction} and \ref{Eq:force_correction} need to be applied.

\subsection{Simulating Electrified Surfaces with Potential Control}
\label{sec:electrochemistry}
\begin{figure} \centering 
\begin{subfigure}[b]{0.55\textwidth} \caption{} 
\includegraphics[width=\textwidth]{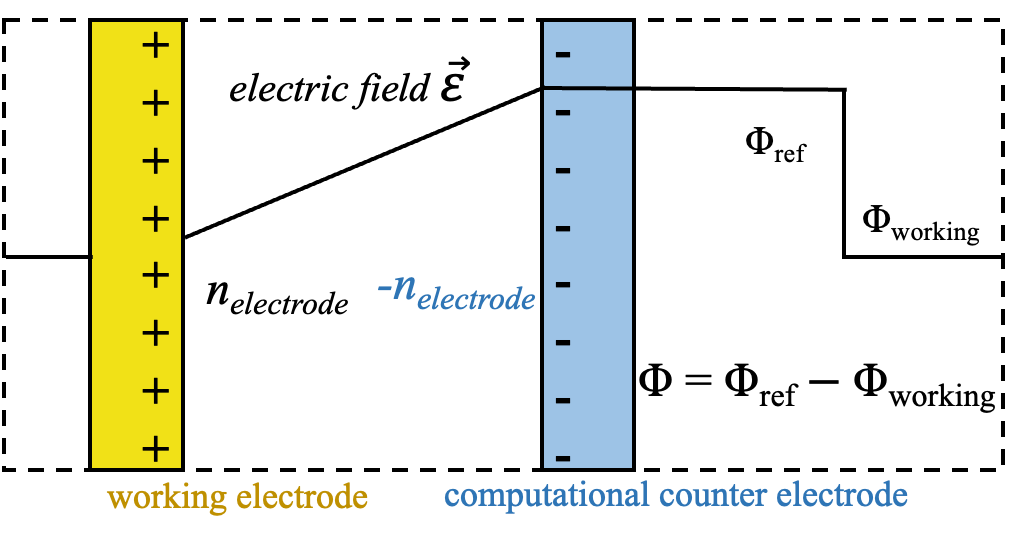} \label{fig:Au111_FSE} \end{subfigure} \begin{subfigure}[b]{0.55\textwidth} \caption{} 
\includegraphics[width=\textwidth]{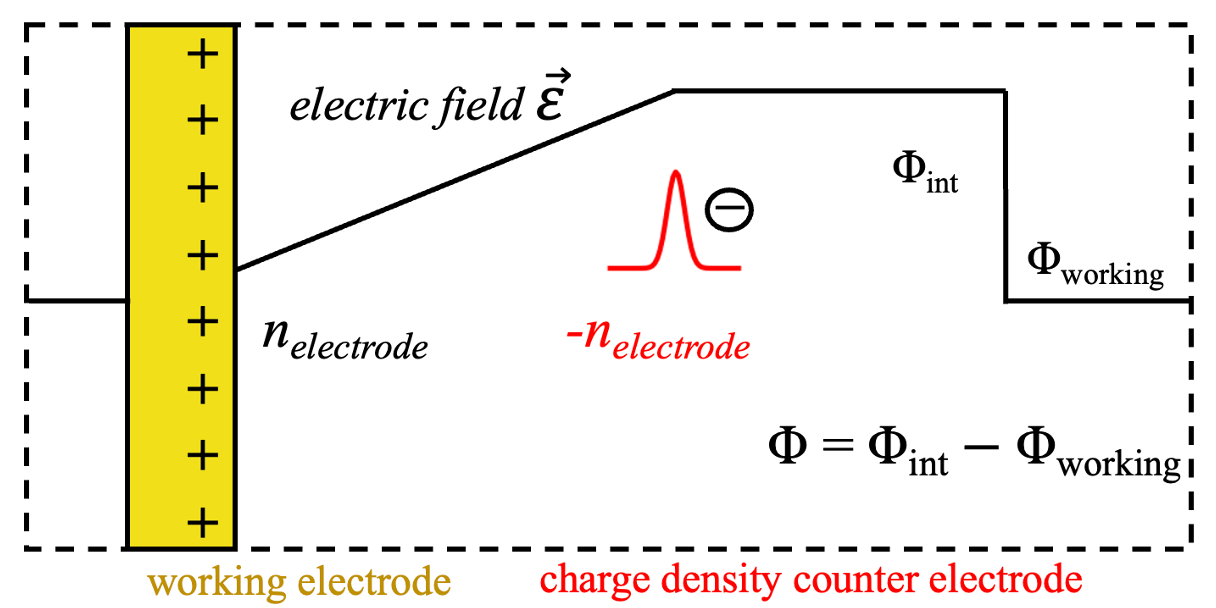} \label{fig:Eads} \end{subfigure} 

\caption{Schematics of simulating charged surface using the a) Ne computational counter electrode (CCE) setup and the b) charge density counter electrode (CDCE) setup. The working electrode has a net charge of $n_{electrode}$, which is compensated by the counter electrode, thus creating an electric field $\mathcal{E}$ across the simulation cell. The black lines represent the electrostatic potential $\Phi$. } \label{fig:CE_schematic} \end{figure}

An important application of field-dependent DFT calculations is to simulate electrified surfaces and interfaces.\cite{todorova2024principlesapproachesconceptselectrochemical} This is achieved by setting up a counter electrode which compensates the charge on the electrified model surface, thus inducing an electric field across the slab system. In previous works, a neon (Ne) computational counter electrode (CCE) has been developed and used, where the counter electrode charge can be adjusted by changing the nuclear core charge of Ne.\cite{Surendralal2018-ug, Deisenbeck2021-lb} The CCE model setup is shown schematically in Fig. \ref{fig:CE_schematic}a. We denote the net charge on the working electrode as $n_{electrode}$. In the CCE setup, the Ne counter electrode carries a net charge of $-n_{electrode}$ and therefore the cell is charge neutral. This configuration allows us to induce a tunable electric field $\mathcal{E}^{\mathrm{ext}}$. The magnitude of the field can be controlled by the position of the CCE and/or the charge $n_{electrode}$ on the electrode.

In this work, we introduce a computational setup beyond CCE, which we term as the charge density counter electrode (CDCE)\cite{Fu1989, Lozovoi2003}. In the CDCE scheme, instead of having explicit atoms for the counter electrode, we introduce an external counter charge density, $\rho_{ext}$, with total charge equal to $-n_{electrode}$. $\rho_{ext}$ is constructed as a sheet charge with a Gaussian distribution in the $z$-direction. In practice, we calculate the corresponding $V_{ext}$ of $\rho_{ext}$ by solving the Poisson equation with fast Fourier transformation (FFT)
\begin{equation}
    \nabla^2 V_{ext}(\mathbf{r}) = -\frac{\rho_{ext}(\mathbf{r})}{\varepsilon_0}\quad.
\end{equation}
$V_{ext}$ is then added to the total potential through the VASP-Python plugin. As such, the $-n_{electrode}$ charge is not accounted for in the DFT self-consistency loop. The total number of electrons in the system is then set to 
\begin{equation}
    n_{total} = \Sigma_I Z_I + n_{electrode}\quad.
    \label{eq:ntotal}
\end{equation}
Here $\Sigma_I Z_I$ sums over the nuclear core charges of all atoms in the system, and is equal to the total number of electrons in a charge-neutral cell. This setup establishes the same $\mathcal{E}^{\mathrm{ext}}$ across the cell as CCE (Fig. \ref{fig:CE_schematic}b). 

The major advantage of going from CCE to CDCE is to increase the range of electric field strength that can be applied. In the CCE method, the maximum field strength applicable is constrained by the Ne band gap. When a sufficiently large field is applied across the cell, the Fermi level of the system goes below the valence band maximum or above the conduction band minimum of Ne, resulting in dielectric breakthrough. This constraint is eliminated when the Ne atoms are replaced by a Gaussian charge density. However, we note that other constraints exist for the maximum field strength, for example the band gap of the electrolyte and the alignment of the Fermi level with respect to the vacuum level. A more detailed discussion can be found in Ref. \citenum{todorova2024principlesapproachesconceptselectrochemical}.

Electrochemical experiments are routinely carried out under constant electrode potential. To correctly represent the experimental setup in AIMD simulations, it is essential to define realistic electrostatic boundary conditions. For a potentiostat, this includes the thermal fluctuations in the electrode charge. One such approach is the thermopotentiostat technique proposed by Dei\ss enbeck\cite{Deisenbeck2021-lb,Deisenbeck2023-jb}. The key ingredient of the method is to introduce the fluctuation term in the electrode charge $n_{electrode}$, to mimic the macroscopic constant-potential dynamics within the small DFT cell. Given an electrode charge at a certain timestep $n_{electrode}(t)$, the electrode charge at the next timestep, $n_{electrode}(t+\Delta t) $ is given by
\begin{equation}
n_{electrode}(t+\Delta t) = n_{electrode}(t) - C_0[\Phi(t) - \Phi_0]\left(1 - e^{-\Delta t / \tau_\Phi}\right) 
 +N\sqrt{k_B T C_0 \left( 1 - e^{-2\Delta t / \tau_\Phi} \right) }\quad,
\label{Eq:thermopotentiostat}
\end{equation}
where $C_0$ is the bare capacitance of the electrodes in vacuum, $\Phi(t)$ is the instantaneous potential measured at the dipole correction, $\Phi_0$ is the target potential, $\tau_\Phi$ is the relaxation time constant, $T$ is the temperature and $k_B $ is Boltzmann's constant. $N$ is a random number with zero mean and variance one. Equation \ref{Eq:thermopotentiostat} essentially sets $n_{electrode}$ at every timestep to achieve $\Phi_0$ on average, with the consistent fluctuations according to statistical physics.

In Ref. \citenum{Deisenbeck2023-jb}, the thermopotentiostat calculations were performed by directly modifying the VASP source code. In this work, we reimplement the method using the VASP-Python interface to enable such simulations with the standard VASP version. Additionally, we have extended the thermopotentiostat method based on the Ne CCE to the CDCE as well. 

\subsection{Implementation in VASP}
In this section, we provide a practical guide on how to use the VASP Python plugin to apply an arbitrary electric field in a simulation. This is done by creating an additional input file for a VASP calculation named \texttt{vasp\_plugin.py}. This Python file provides callback functions that can access certain quantities during the simulation and allow the user to modify some of them. For the purpose of this work, we use three of these functions. 

\begin{lstlisting}[style=pyclean]
def local_potential(constants, additions): 
# adds the external field and correct the energies
def force_and_stress(constants, additions): 
# corrects the forces
def occupancies(constants, additions): 
# changes the applied bias at each ionic step
\end{lstlisting}

In each of these functions, \texttt{constants} provides a dataclass that contains given calculated quantities during the VASP run. These quantities are accessible to the user, but cannot be changed. \texttt{additions}, on the other hand, contains quantities that can be changed by the user. The list of quantities included in each dataclass can be found in the VASP documentation. For example, in the \texttt{local\_potential} function, we have constants
\begin{lstlisting}[style=pyclean]
class ConstantsLocalPotential:
    ENCUT: float
    NELECT: float
    ...
    charge_density: Optional[DoubleArray] = None
    hartree_potential: Optional[DoubleArray] = None
    ...
\end{lstlisting}
\begin{lstlisting}[style=pyclean]
class AdditionsLocalPotential:
    total_energy: float
    total_potential: DoubleArray

\end{lstlisting}
So in the \texttt{local\_potential} function in \texttt{vasp\_plugin.py}, we can access the energy cutoff (ENCUT) with \texttt{constants.ENCUT} (similarly for all other accessible quantities), and we can change the total potential with
\begin{lstlisting}[style=pyclean]
additions.total_potential += delta_potential
\end{lstlisting}

To call these functions defined in \texttt{vasp\_plugin.py}, one also needs to set the corresponding flags to true in the INCAR file:

\begin{lstlisting}[style=pyclean]
PLUGINS/LOCAL_POTENTIAL = T
PLUGINS/FORCE_AND_STRESS = T
PLUGINS/OCCUPANCIES = T
\end{lstlisting}    
In the following sections, we briefly explain the role of each function and discuss related technical details. 

\subsubsection{\texttt{local\_potential}}

The task of the \texttt{local\_potential} routine is to modify the electrostatic potential that VASP uses internally.  
The modification is performed by adding an external contribution \(V_{\text{ext}}\) to the existing potential:
\begin{lstlisting}[style=pyclean]
additions.total_potential += delta_potential
\end{lstlisting}

Here \texttt{delta\_potential} is a three‑dimensional array that must have the same mesh as VASP’s electrostatic potential and charge density (dimensions \(\text{NGX}\times\text{NGY}\times\text{NGZ}\)). The \texttt{local\_potential} function has access to the charge density \texttt{constants.charge\_density} and the dimensions can be directly obtained.  

In addition to updating the potential, two further corrections are required. First, VASP does {\em not} include the interaction of the external potential with the nuclear charges, so the corresponding energy term has to be added explicitly:
\begin{lstlisting}[style=pyclean]
additions.total_energy += delta_energy
\end{lstlisting}
The quantity \texttt{delta\_energy} is defined as  
\[
\Delta E = \sum_{I}\bigl[-Z_{I}\,V_{\text{ext}}(\mathbf{R}_{I})\bigr],
\]
where the sum runs over all nuclei \(I\) with charge \(Z_{I}\) located at \(\mathbf{R}_{I}\).

Second, special care is needed when a dipole correction is required.  
Consider the CDCE configuration illustrated in Fig.~\ref{fig:CE_schematic}b. In this setup \(V_{\text{ext}}\) represents the field generated by the counter‑electrode charge \(-n_{\text{electrode}}\).  
The total number of electrons that VASP should treat is  
\[
n_{\text{tot}} = n_{\text{neutral}} + n_{\text{electrode}},
\]
with \(n_{\text{neutral}}\) being the electron count of a neutral cell.  

Because VASP does {\em not} contain the charge density of the counter electrode, it interprets the system as non‑neutral. If the dipole correction is activated in VASP (via the tag \texttt{LDIPOL = .TRUE.}), the program aborts, since the dipole correction is only implemented for charge‑neutral cells.
To circumvent this limitation we disable VASP’s built‑in dipole correction and implement it manually inside the \texttt{local\_potential} routine. The contribution of the counter‑electrode charge is computed in the helper function \texttt{calc\_dipole} and added to \texttt{delta\_potential} before the potential update:
\begin{lstlisting}[style=pyclean]
rho_dipole = calc_dipole(rho+rho_external)
delta_potential += electrostatic_potential(rho_dipole)
additions.total_potential += delta_potential
\end{lstlisting}
Here \texttt{rho} is the charge density in VASP (excluding the counter electrode charge density), \texttt{rho\_external} is the charge density of the externally-added counter electrode. The \texttt{calc\_dipole} function then computes the total dipole of the system and outputs the dipole charge density that should be added in the vacuum region \texttt{rho\_dipole}. The \texttt{electrostatic\_potential} function then computes the corresponding potential profile of the charge density, which is added to the total potential. By doing so the dipole correction is applied consistently, even for simulations that involve a net charge introduced by the external electrode.

\subsubsection{\texttt{force\_and\_stress}}

The \texttt{force\_and\_stress} routine implements the force correction that appears in Eq.~\ref{Eq:force_correction}, i.e. the force exerted on each nucleus by the externally applied field.  
For the CDCE configuration a short‑range repulsive “wall’’ is also added near the counter‑electrode to keep the water molecules confined; this wall force is obtained from the helper function \texttt{wall}.  
Both contributions are combined and added to the forces that VASP returns:

\begin{lstlisting}[style=pyclean]
delta_force = force_wall + force_ion
additions.forces += delta_force
\end{lstlisting}
In this work, we consider only calculations with fixed cell size, so the stress contributions are not considered.

\subsubsection{\texttt{occupancies}}

The \texttt{occupancies} routine is called at the end of every ionic step.  It enables a time‑dependent bias during a molecular‑dynamics run, as required by the thermopotentiostat scheme.  In the charge‑density counter‑electrode setup (Fig.~\ref{fig:CE_schematic}b) the routine performs two tasks.

\paragraph{(i) Update of the electrode charge}  
The change in the electrode charge, \(\Delta n_{\text{electrode}}\equiv dq\), for the next step is evaluated from Eq.~\ref{Eq:thermopotentiostat}:

\begin{lstlisting}[style=pyclean]
dq = C0*(phi0 + (phi - phi0)*np.exp(-1.0/tau) \
      + np.sqrt(kB * temperature / C0) \
      * np.sqrt(1.0 - np.exp(-2.0/tau)) \
      * np.random.normal() - phi)
\end{lstlisting}

Here \texttt{C0} is the system capacitance, \texttt{phi0} the target bias, and \texttt{phi} the instantaneous bias computed at the current step.  The value of \texttt{dq} is stored so that the \texttt{local\_potential} function can read it and update the external potential \(V_{\text{ext}}\).

\paragraph{(ii) Adjustment of the total electron count}  
Because the electrode charge changes, the total number of electrons in the simulation must be modified in accordance with Eq.~\ref{eq:ntotal}.  This is done by updating the \texttt{NELECT} variable:

\begin{lstlisting}[style=pyclean]
additions.NELECT -= dq
\end{lstlisting}

\subsubsection{Additional development features}

Two capabilities that were employed in this work are available only in the development branch of VASP and have not yet been released in the official distribution:

\begin{enumerate}
  \item \textbf{Ne computational counter electrode.}  
    Instead of modifying \(V_{\text{ext}}\) and \texttt{NELECT}, the number of core electrons of neon (\texttt{ZVAL}) is changed.  In VASP 6.5.0 the \texttt{ZVAL} field is not part of the \texttt{additions} dataclass used by \texttt{occupancies}, but this will be added in a forthcoming release.

  \item \textbf{Force‑drift correction.}  
    By default VASP applies a drift‑correction that forces the net force on all atoms to vanish.  When an external potential is present this constraint is no longer appropriate.  In the development version the drift correction can be disabled with the flag  

\begin{lstlisting}[style=pyclean]
LREMOVE_DRIFT = .FALSE.
\end{lstlisting}
    This option will also become available in the next official VASP release.
\end{enumerate}





\section{Case Studies}

By utilizing Python scripting within VASP, we are able to apply an arbitrary external electric field in a DFT calculation. In this section, we present four case studies utilizing the implemented method: (1) adsorption on electrified surfaces; (2) field ion microscopy; (3) electrochemical interfaces; and (4) implicit solvation model. These examples highlight the range of field-induced physical and chemical processes that can be modeled with DFT and the flexibility and robustness of our implementation.  

\subsection{Adsorption on electrified surfaces}

\begin{figure}
    \centering
    \begin{subfigure}[b]{0.45\textwidth}
        \caption{}
        \includegraphics[width=\textwidth]{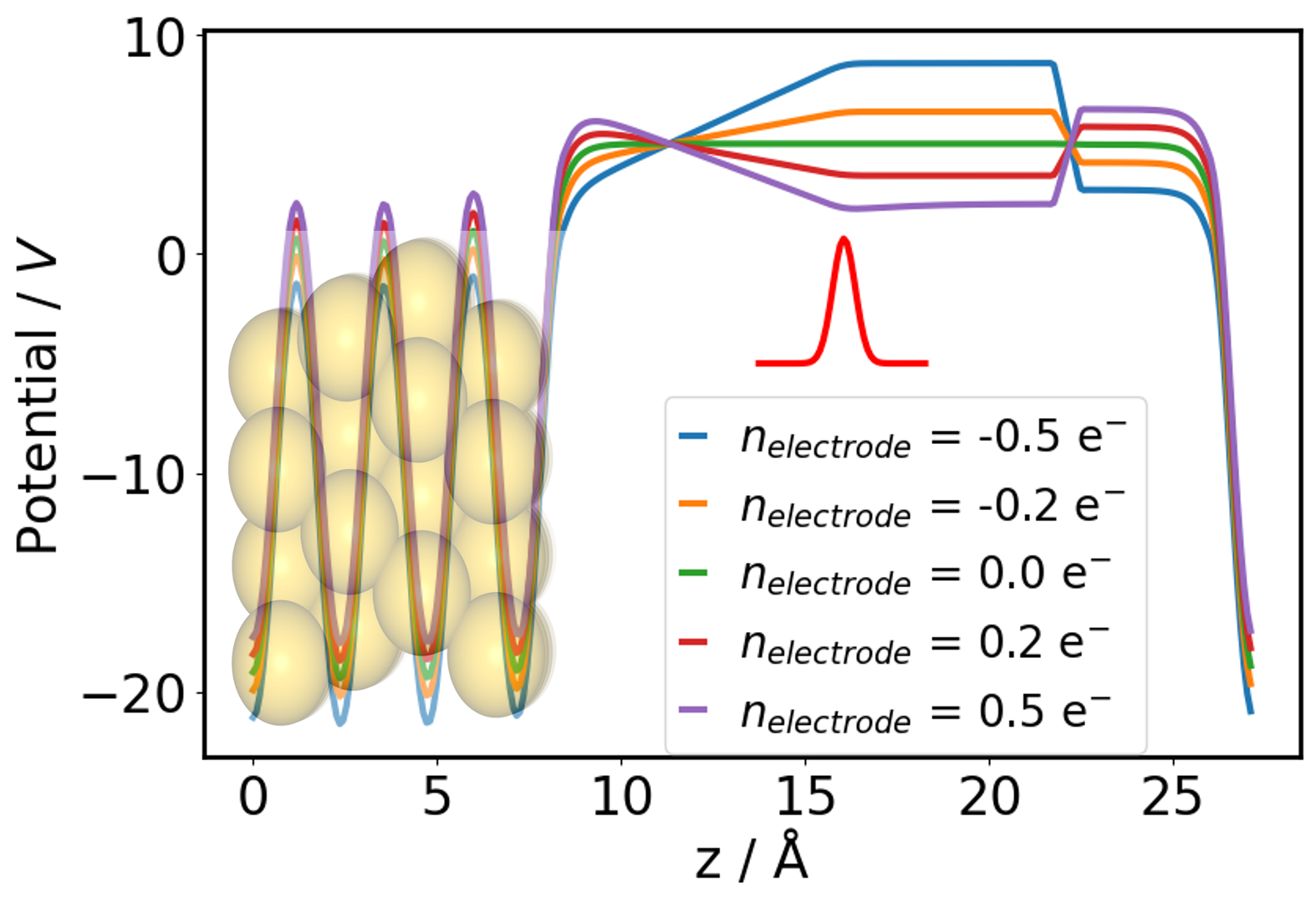}
        \label{fig:Au111_FSE}
    \end{subfigure}
    \begin{subfigure}[b]{0.45\textwidth}
        \caption{}
        \includegraphics[width=\textwidth]{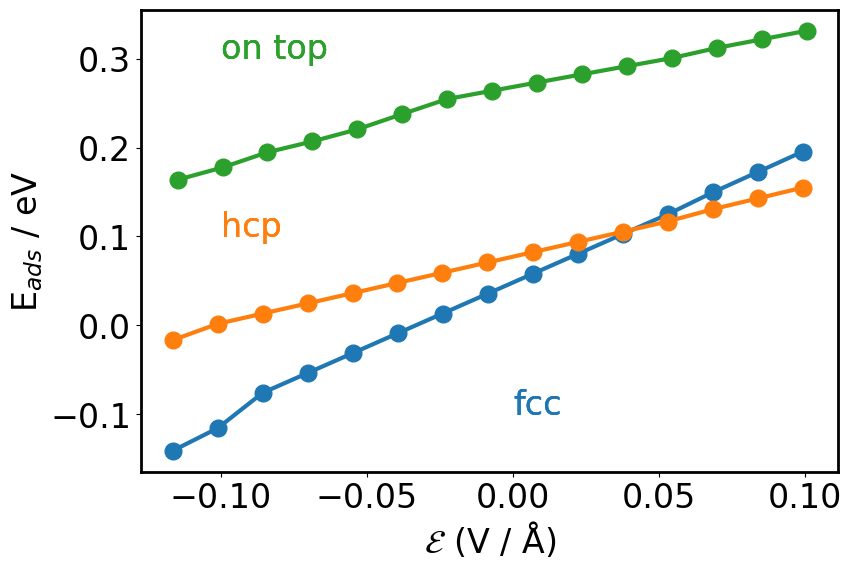}
        \label{fig:Eads}
    \end{subfigure}
    \begin{subfigure}[b]{0.4\textwidth}
        \caption{}
        \includegraphics[width=\textwidth]{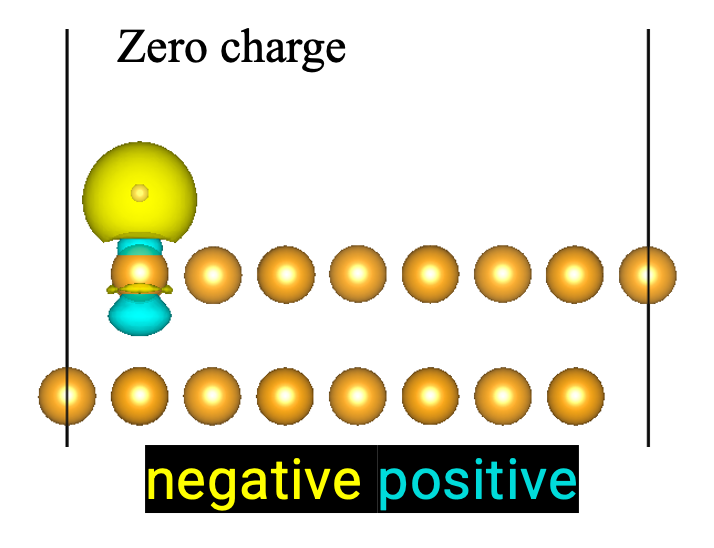}
        \label{fig:charge_density_diff}
    \end{subfigure}
    \begin{subfigure}[b]{0.4\textwidth}
        \caption{}
        \includegraphics[width=\textwidth]{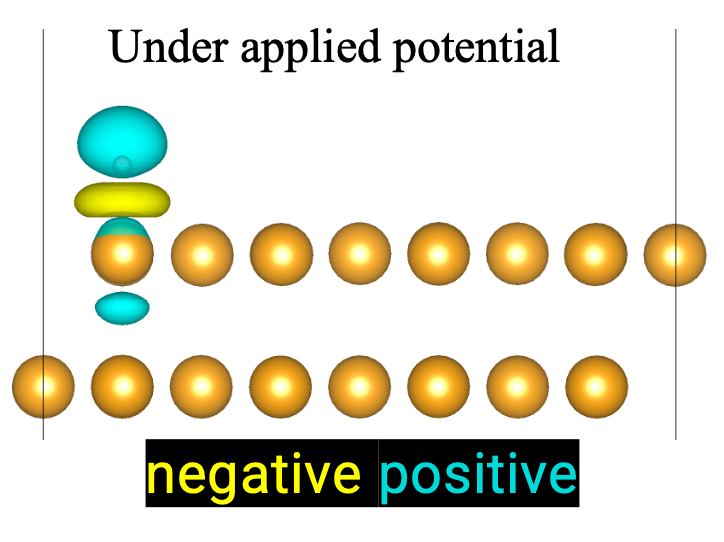}
        \label{fig:charge}
    \end{subfigure}
    \caption{
        a) Electrostatic potential of the Au(111) surface for several electrode charges; the red curve corresponds to the Gaussian charge‑density counter electrode (CDCE).  
        b) Adsorption energies of a single H atom on Au(111) as a function of the applied electric field.  
        c) Charge‑density difference for $n_{\text{electrode}} = 0$, computed with Eq.~\ref{Eq:AuH_chg_diff} and an isosurface value of $1.5\times10^{-6}\ e/\mathrm{\AA}^3$.  
        d) Charge‑density difference for $n_{\text{electrode}} = -0.5\,e^{-}$, computed with Eq.~\ref{Eq:AuH_double_chg_diff} and an isosurface value of $1.5\times10^{-7}\ e/\mathrm{\AA}^3$.  
        Yellow denotes excess electronic charge (negative charge density), while cyan denotes a deficit of electronic charge (positive charge density). 
    }
    \label{fig:Au111_H}
\end{figure}

To demonstrate the capability of our workflow to treat electrified interfaces, we use the Au(111) surface as a benchmark system and impose a series of electrode charges $n_{\text{electrode}}$.  Each chosen $n_{\text{electrode}}$ generates a corresponding electrostatic bias (or voltage) $\Phi$, which can be obtained directly from the macroscopic dipole moment of the slab
\begin{equation}
    \Phi = \frac{\mu_{\mathrm{dip}}}{\epsilon_{0} A}\,,
    \label{Eq:voltage_dipole}
\end{equation}
where $\mu_{\mathrm{dip}}$ is the dipole moment of the simulation cell, $\epsilon_{0}$ the vacuum permittivity, and $A$ the surface area of the cell.  The laterally averaged electrostatic potentials resulting from the Au(111) calculations are shown in Fig.~\ref{fig:Au111_H}a; the red Gaussian‑shaped curve represents the CDCE.  For $n_{\text{electrode}} = 0.5\,e^{-}$ a voltage of $\Phi = 5.8\ \text{V}$ is induced in the Au(111) slab.  This approach for applying a bias to a DFT calculation is readily transferable to any slab system: the position, shape, and magnitude of the external charge density $\rho_{\text{ext}}$ (and thus $n_{\text{electrode}}$) can be tuned, provided the resulting field does not exceed the dielectric breakdown limit \cite{Yoo2021,Freysoldt2018}.

Identifying active sites and reaction mechanisms is central to the development of electrocatalysts.  Surface adsorption under an applied potential reveals the thermodynamics of the interface at a given electrochemical condition and is therefore key to uncovering reaction pathways \cite{Govindarajan2025-hv}.  As a showcase, we examine hydrogen adsorption on Au(111) using the CDCE approach.  With Au(111) as the reference surface and a single H atom as the adsorbate, the adsorption energy under a bias $\Phi$ is evaluated as
\begin{equation}
    E^{\text{Au/H}}_{\text{ads}}(\Phi) = 
    E^{\text{Au/H}}_{\text{tot}}(\Phi) 
    - E^{\text{Au}}_{\text{tot}}(\Phi) 
    - \frac{E^{\text{H}_{2}}_{\text{tot}}}{2}\,,
    \label{Eq:AuH_Eads}
\end{equation}
where $E^{\text{Au/H}}_{\text{tot}}(\Phi)$ and $E^{\text{Au}}_{\text{tot}}(\Phi)$ are the total energies of the slab with and without the adsorbate, respectively, and $E^{\text{H}_{2}}_{\text{tot}}$ is the total energy of an isolated H$_2$ molecule.  The resulting hydrogen adsorption energies as a function of the applied field are displayed in Fig.~\ref{fig:Au111_H}b.  Three adsorption sites are considered: on‑top, hcp‑hollow, and fcc‑hollow.  At zero field the fcc‑hollow site is the most stable, but the hcp‑hollow site becomes energetically favored over the fcc‑hollow site at an electric field of approximately $0.05\ \text{V\,\AA}^{-1}$.


The strength of these adsorbate-surface bonds is determined by the amount of electron transfer, which controls the kinetics of the adsorption and desorption of intermediates during the catalytic cycle\cite{Liao2022-br}. Electron redistribution during the catalytic cycle is therefore critical to understanding catalytic performance \cite{Chen2019-cv}. 
To visualize and analyze these charge transfer processes, one can calculate the charge density difference between the adsorbate on the surface and the bare surface and isolated species. With the CDCE methodology, this is also possible under applied potential, where the electronic charge density difference is given by 

\begin{equation}
\Delta\rho_1 = \rho^{Au(111)/H}_{n_{electrode}} - \rho^{Au(111)}_{n_{electrode}}-
\rho^{H}_{ZC},
\label{Eq:AuH_chg_diff}
\end{equation}
with zc representing zero electrode charge, i.e., no external applied potential. 

Figure \ref{fig:Au111_H}c illustrates this charge density difference, which is dominated by the charge transfer of the surface to the H atom irrespective of $\Phi$, meaning it is difficult to visually differentiate between the charge density difference calculated by Equation \ref{Eq:AuH_chg_diff} when $n_{electrode} = -0.5 e^{-}$ or $n_{electrode} = 0$. 
Therefore, to isolate the charge transfer occuring upon applying an external bias from the charge transfer of the Au-H bond formation, we calculate the double charge density difference, given by 
\begin{equation}
\Delta\rho_2 = (\rho^{Au(111)/H}_{n_{electrode}} - 
\rho^{Au(111)/H}_{zc} )- 
(\rho^{Au(111)}_{n_{electrode}} - 
\rho^{Au(111)}_{zc} )
\label{Eq:AuH_double_chg_diff}
\end{equation}
and shown in Figure \ref{fig:Au111_H}d. Here we visualize the polarization the Au-H bond under the effect of $\Phi$. To quantify the charge transfer of Equations \ref{Eq:AuH_chg_diff} and \ref{Eq:AuH_double_chg_diff}, the charge density differences can be integrated as

\begin{equation}
\Delta n_1 = \frac{1}{2}\int_{}\hspace{7pt}
\Bigl|
\bigl(
   \rho^{\mathrm{Au(111)/H}}_{\,n_{\text{electrode}}}
   -\rho^{\mathrm{Au(111)}}_{\,n_{\text{electrode}}}
   -\rho^{\mathrm{H}}_{\text{zc}}
\bigl)
\Bigr|
\,\mathrm{d}V,
\label{eq:integrated_abs_charge_diff}
\end{equation}

\begin{equation}
\Delta n_2 = \frac{1}{2}
{
\int_{}\hspace{7pt}
\Bigl|
\bigl(\rho^{\mathrm{Au(111)/H}}_{\,n_{\text{electrode}}}
      -\rho^{\mathrm{Au(111)/H}}_{\text{zc}}\bigr)
-
\bigl(\rho^{\mathrm{Au(111)}}_{\,n_{\text{electrode}}}
      -\rho^{\mathrm{Au(111)}}_{\text{zc}}\bigr)
\Bigr|
\,dV.
}
\label{eq:integrated_abs_charge_diff2}
\end{equation}
Equation \ref{eq:integrated_abs_charge_diff} gives a charge of 1.29 and 1.28$e^{-}$ for $n_{electrode}$ equal to zero and $-$0.5$e^{-}$, respectively, while Equation \ref{eq:integrated_abs_charge_diff2} gives a charge
 0.09 $e^{-}$ for $n_{electrode}$ = $-$0.5$e^{-}$. This shows how the formation of the interaction of H with the Au dominates the charge transfer process; an additionally potential bias then only slightly further polarizes the surface-adsorbate. The positively charged gold surface polarizes the electron density along the Au-H bond towards the surface.

\begin{figure}
  \centering
  \begin{subfigure}[b]{0.65\textwidth}
    \includegraphics[width=\textwidth]{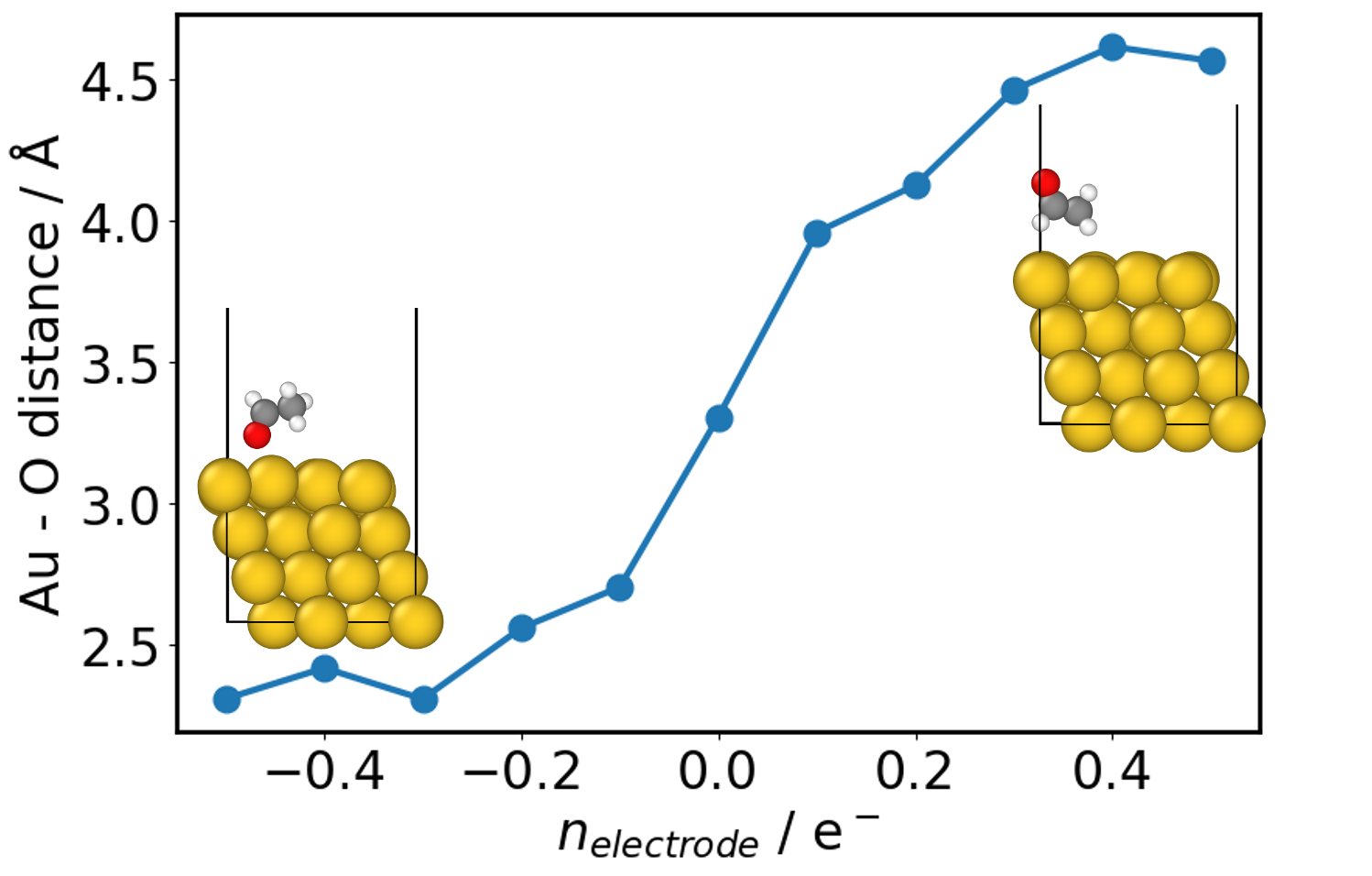}
  \end{subfigure}
  \hfill
    \caption{AIMD simulations of an acetaldehyde molecule adsorbed on the Au(111) surface under applied field, showing the dependence of Au-O distance with $n_{electrode}$. }
  \label{fig:Aldehyde_MD}
\end{figure}

The CDCE methodology can be additionally used for AIMD, enabling a wide range of investigations. As test system, we perform AIMD simulations to probe the interaction of an acetaldehyde molecule with the Au(111) surface under a range of electrode charges (Fig. \ref{fig:Aldehyde_MD}). The finite temperature simulation allows us to explore the diverse configurations of the organic molecule under applied field. The positively polarized surface, i.e. $n_{electrode}$ $<$ 0, interacts strongly with the electronegative oxygen of the aldehyde, resulting in a small Au-O distance. As $n_{electrode}$ increases and the surface becomes negatively polarized, the oxygen is repelled. This results in the rotation of the acetaldehyde molecule with changing electrode charge. These simulations illustrate how our methodology can be used to explore dynamic systems under applied field, leading to possible mechanistic insights on reactions or interface processes. 

\subsection{Atom probe tomography}

\begin{figure}
    \centering
    \begin{subfigure}[b]{0.45\textwidth}
        \caption{}
        \includegraphics[width=\textwidth]{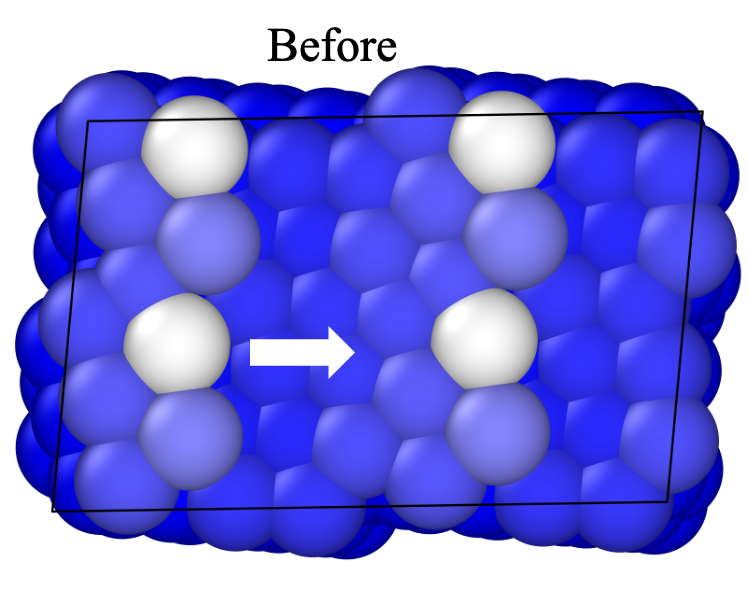}
    \end{subfigure}
    \begin{subfigure}[b]{0.45\textwidth}
        \caption{}
        \includegraphics[width=\textwidth]{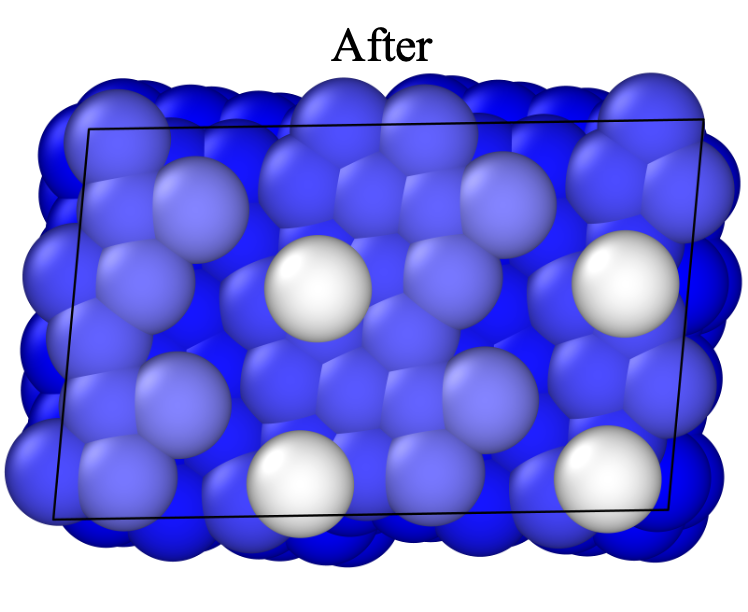}
    \end{subfigure}
    \caption{
        a) Initial and b) final snapshot of a 5‑ps \textit{ab‑initio} molecular‑dynamics (AIMD) trajectory of the (952) kinked Li surface at 300~K under an applied electric field of \(1.25~\mathrm{V\,\AA^{-1}}\).  
        The atom coloured white is the kink atom, which is emitted to an adatom position after \(\approx 4.5\)~ps of simulation time.  
        Lighter atoms correspond to surface atoms.}
    \label{fig:Li_field}
\end{figure}

Modeling electrified surfaces is of central importance for field ion microscopy (FIM) and atom probe tomography (APT), where surface atoms undergo field‑induced desorption, ionisation, and evaporation.  Recent work on the Li(110) surface demonstrated that a strong electric field can alter the preferred adsorption site of Li adatoms and even enable barrier‑less surface diffusion \cite{Katnagallu2025-aq}.  The same phenomenon can be captured directly in AIMD simulations that employ the CDCE setup.

Here we simulated a kinked Li(952) surface at 300~K.  The top‑view of the initial structure is shown in Fig.~\ref{fig:Li_field}a, with the kink atom highlighted in white.  In the absence of an external field the kink atom remains bound to the step edge and no diffusion events occur.  When a uniform field of \(1.25~\mathrm{V\,\AA^{-1}}\) is applied, the kink atom spontaneously detaches and migrates across the surface within \(\sim 4.5\)~ps of AIMD time, as illustrated in Fig.~\ref{fig:Li_field}b.

These calculations provide a direct, atomistic confirmation that strong electric fields can destabilise kink/step atoms and promote their release as adatoms at finite temperature.  The observed behaviour corroborates thermodynamic analyses predicting the destabilisation of surface adatom states under high fields\cite{Katnagallu2025-aq}, and offers valuable insight into the mechanisms governing APT experiments.


    

\subsection{Electrochemical interfaces}

\begin{figure}
    \centering
    \begin{subfigure}[b]{0.65\textwidth}
        \caption{}
        \includegraphics[width=\textwidth]{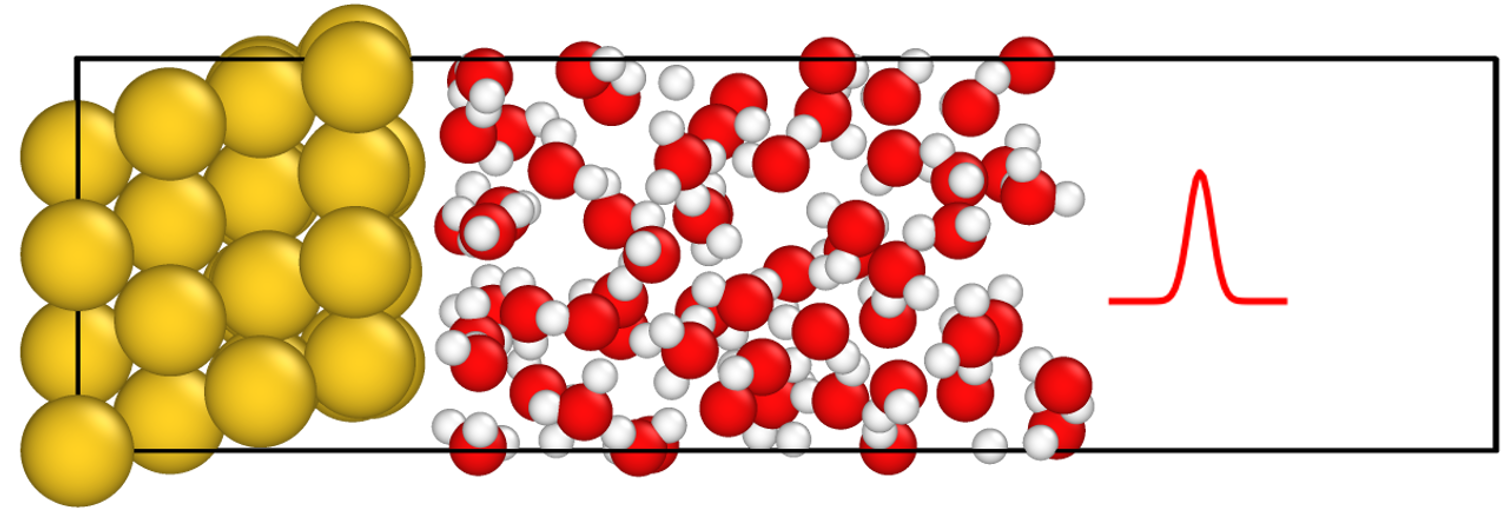}
        \label{fig:Au111_FSE}
    \end{subfigure}
    \begin{subfigure}[b]{0.5\textwidth}
        \caption{}
        \includegraphics[width=\textwidth]{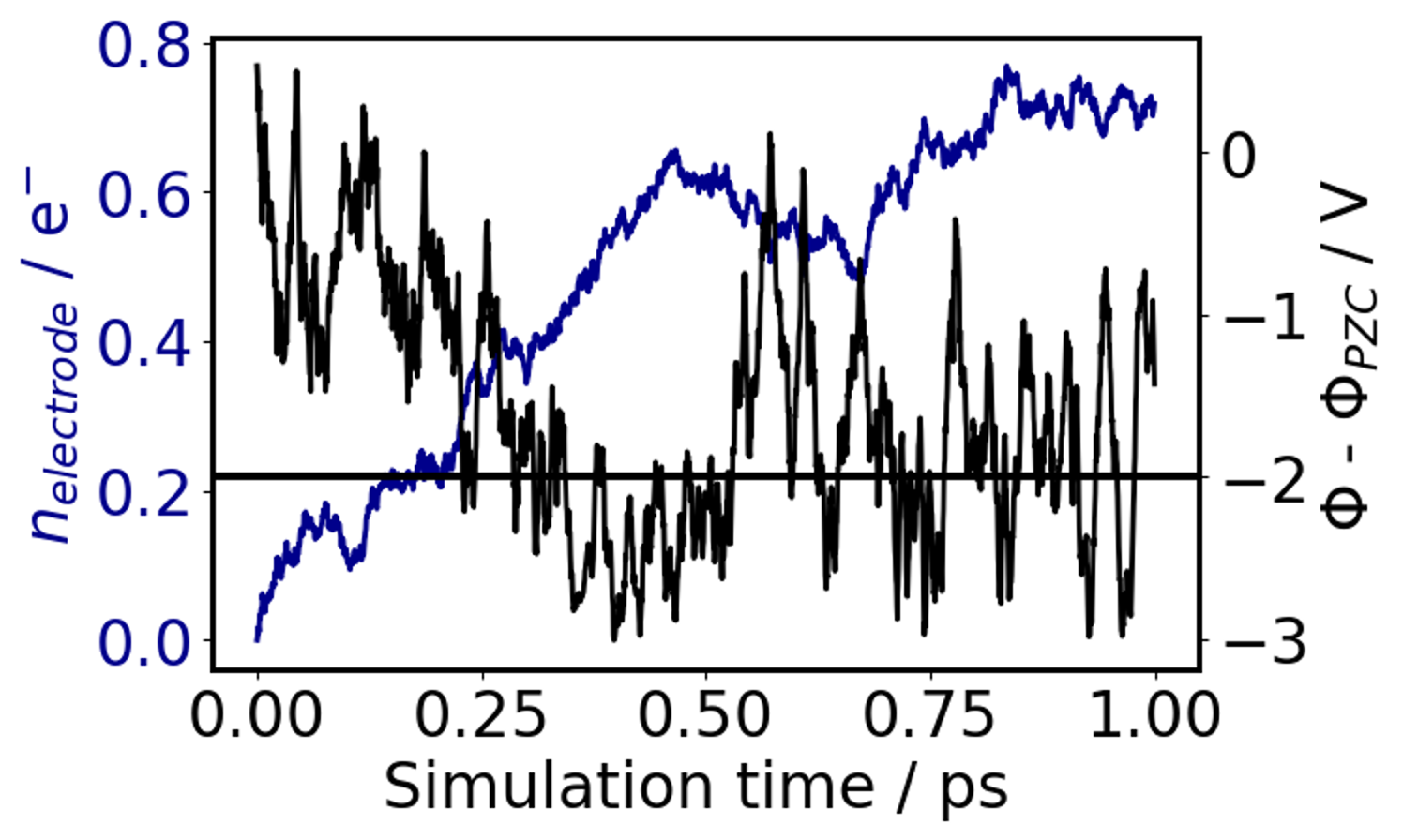}
        \label{fig:Eads}
    \end{subfigure}
    \begin{subfigure}[b]{0.44\textwidth}
        \caption{}
        \includegraphics[width=\textwidth]{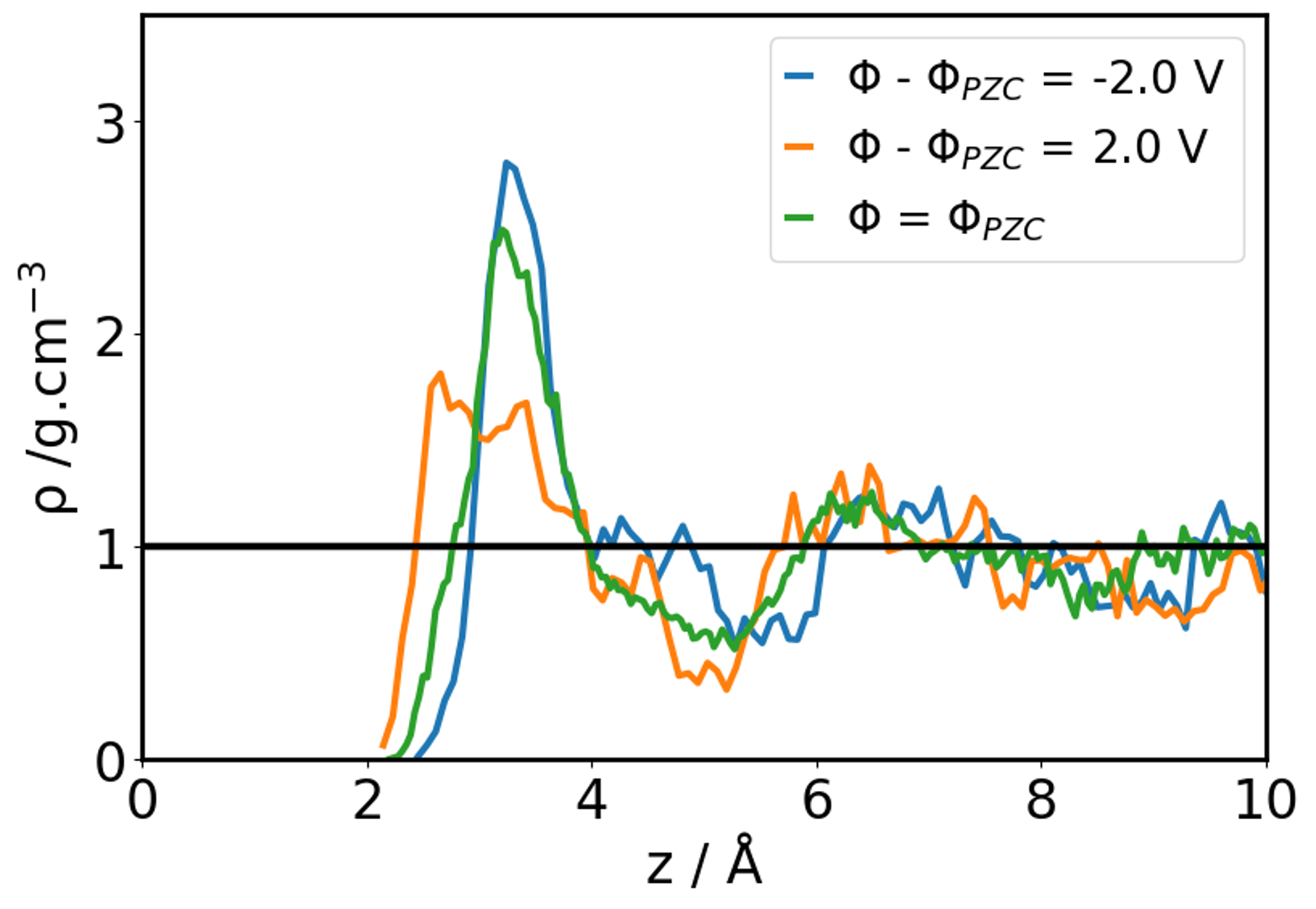}
        \label{fig:Eads_density}
    \end{subfigure}
    \caption{
        a) Au(111)/H$_{2}$O slab with the CDCE setup.  
        b) Time evolution of the electrode charge and the corresponding electrode potential.  The target potential $\Phi_{0}-\Phi_{\mathrm{zc}}$ (horizontal line) is set to $-2.0\,$V.  
        c) Oxygen density profiles obtained from AIMD simulations of Au(111)/H$_{2}$O under different applied voltages and at the potential of zero charge.}
    \label{fig:Au_H2O_everything}
\end{figure}

Thus far we have described how to impose an external electric field on a vacuum‑terminated slab.  Extending the approach to solid–liquid interfaces requires additional care to avoid that the computational counter electrode affects water structure and dynamics.  The Ne computational counter electrode (Ne‑CCE) offers an inert, wide‑band‑gap electrode that has been successfully employed to study Mg corrosion \cite{Surendralal2018-ug,Deisenbeck2024-fk}.  However, its hydrophobic character limits the maximum field that can be sustained across the solid–liquid boundary.  This limitation motivates the development of fully electronic counter electrodes, such as a Gaussian charge density analogous to the CDCE.  A comprehensive discussion of the available approaches can be found in Ref.~\citenum{todorova2024principlesapproachesconceptselectrochemical}.

Figure~\ref{fig:schematic picture}  illustrates the combination of the CDCE with the thermopotentiostat method for the Au(111)/water interface.  During an AIMD run the charge on the gold surface, $n_{\text{electrode}}$, is updated at every time step by the VASP–Python plugin according to Eq.~\ref{Eq:thermopotentiostat}.  This feedback loop enforces a constant electrode potential, enabling truly potential‑controlled AIMD simulations of solid–liquid interfaces.
The resulting dynamics of $n_{\text{electrode}}$ and the instantaneous electrode potential $\Phi$ are displayed in Fig.~\ref{fig:Au_H2O_everything}b.  As the simulation progresses, the thermopotentiostat adjusts $n_{\text{electrode}}$ so that $\Phi$ approaches the prescribed target $\Phi_{0}$.

The applied potential has a pronounced effect on the interfacial water structure, as evident from the oxygen density profiles in Fig.~\ref{fig:Au_H2O_everything}c.  At negative potentials the oxygen atoms are repelled from the gold surface, shifting the first density peak away from the interface.  Conversely, at positive potentials the surface charge attracts oxygen, pulling the two main peaks of the density profile closer to the metal.  These structural changes are accompanied by variations in vibrational spectra, capacitance, and dynamical properties of the water layer \cite{Goldsmith2021-mu,Li2025-zn}, underscoring the importance of explicit potential control in simulations of electrochemical interfaces.

\subsection{QM/MM solvation models }

Beyond explicit solid‑liquid simulations, the ability to modify the external potential $V_{\text{ext}}$ makes it straightforward to incorporate a variety of solvation schemes. In particular, one can replace the explicit solvent by an effective external potential $V^{\text{solvent}}$, enabling implicit solvent models or QM/MM (quantum‑mechanics/molecular‑mechanics) approaches.

\begin{figure} \centering 
\begin{subfigure}[b]{0.48\textwidth} \caption{} 
\includegraphics[width=\textwidth]{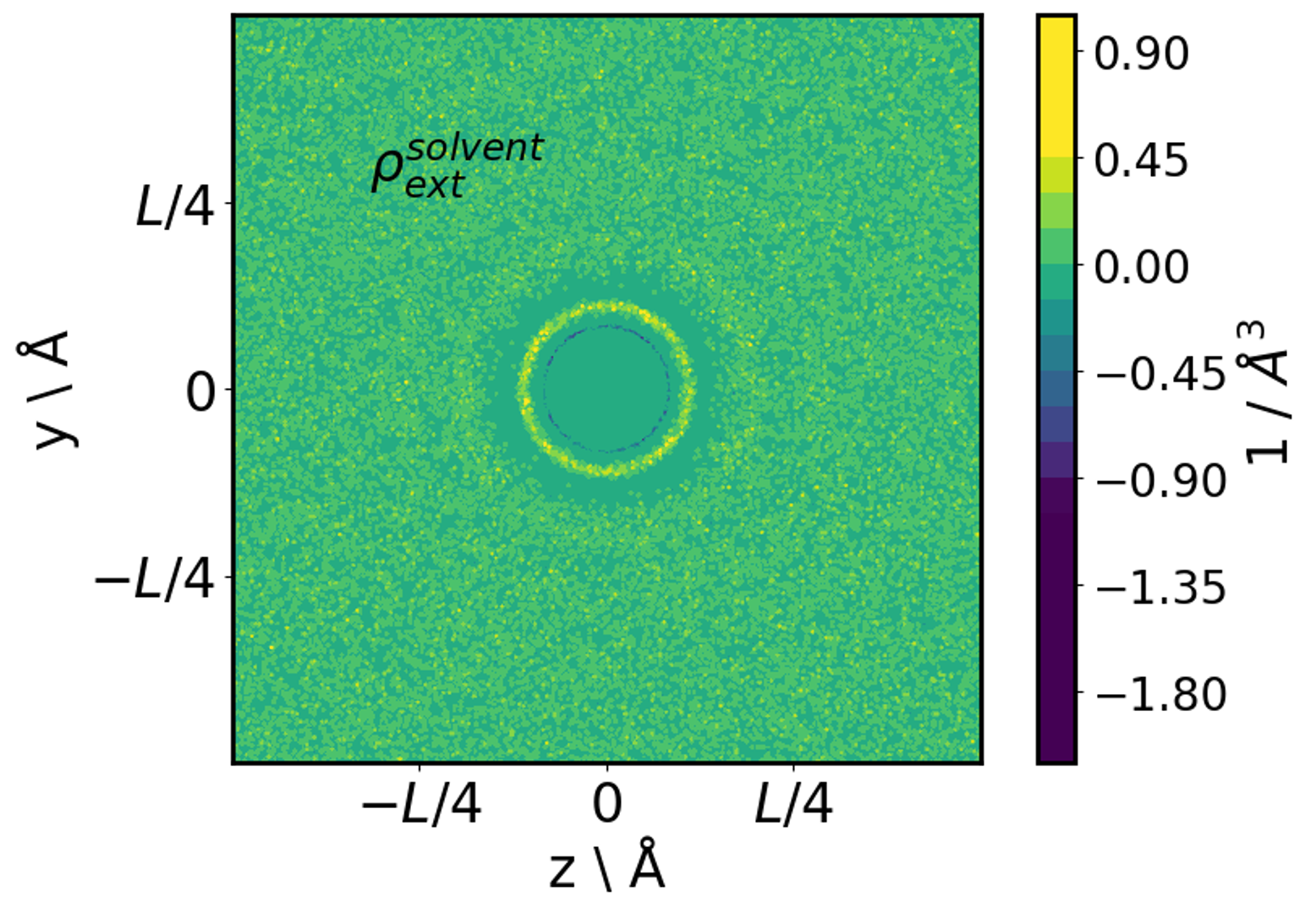} \label{fig:Au111_FSE} \end{subfigure} \begin{subfigure}[b]{0.5\textwidth} \caption{} 
\includegraphics[width=\textwidth]{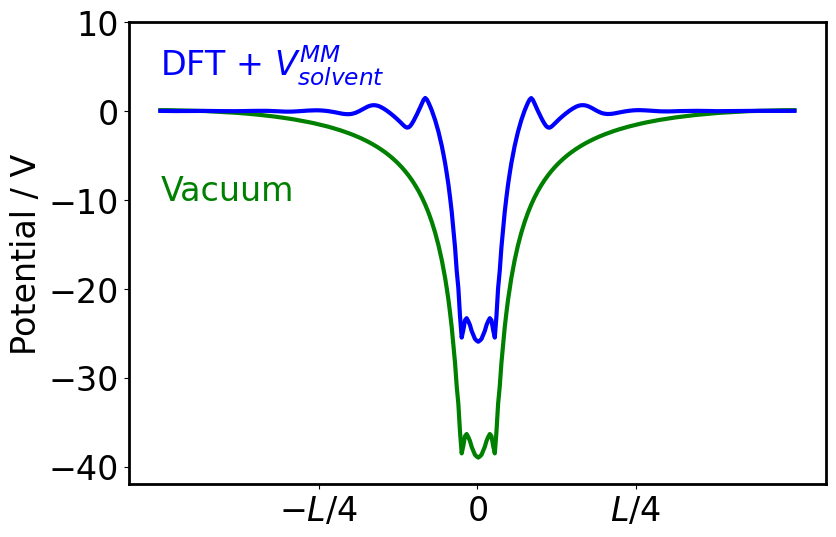} \label{fig:Eads} \end{subfigure} 

\caption{a) Contour plot of the solvent charge density $\rho^{MM}_{solvent}$  of a Mg$^{2+}$ ion solvated in water. b) Comparison between the electrostatic potential of a single Mg$^{2+}$ in vacuum and in the solvated QM/MM method.  
} \label{fig:QM_MM} \end{figure}

Figure~\ref{fig:QM_MM} illustrates a QM/mean-field-MM implicit solvation workflow implemented using the VASP-Python interface. The system under study is a $\mathrm{Mg}^{2+}$ ion dissolved in water. First, a classical molecular‑dynamics (MD) simulation of the aqueous environment is performed for several nanoseconds. During this MD run the charge density of the solvent is averaged, yielding the MM solvent charge density $\rho^{\text{MM}}_{\text{solvent}}$ shown in Fig.~\ref{fig:QM_MM}a. This charge density is then converted into an electrostatic potential $V^{\text{MM}}_{\text{solvent}}$, which is added to the Kohn‑Sham Hamiltonian as the external potential $V_{\text{ext}}$. More details about the QM/MM solvation method will be described in a forthcoming paper.

Figure~\ref{fig:QM_MM}b compares the resulting electrostatic potentials for two cases. In vacuum, the isolated $\mathrm{Mg}^{2+}$ ion exhibits a long‑range, parabolic electrostatic potential (green line). When the solvent electrostatics are included (blue line), the $+2$ charge of the magnesium ion is screened by the surrounding water molecules, producing an oscillatory potential profile that reflects the formation of distinct solvation shells around the ion.

Because the solvent contribution is introduced solely through $V_{\text{ext}}$, the QM/MM scheme can be readily extended to more complex geometries, such as surfaces and interfaces. Users can define arbitrary QM regions and couple them to a flexible MM description of the environment, making the approach highly adaptable to a wide range of electrochemical and interfacial problems.

\section{Conclusion}

In this work we have introduced a flexible scheme for imposing arbitrary electrostatic potentials in density‑functional‑theory (DFT) supercell calculations.  The approach is implemented via the VASP–Python interface, which allows the user to supply any desired external potential \(V_{\text{ext}}\) and to have it incorporated directly into the Kohn–Sham Hamiltonian.  Because the external field also interacts with the ionic cores, we derived and applied the necessary energy‑ and force‑correction terms to ensure that total energies and atomic forces remain physically meaningful.

The utility of the method is illustrated through several representative case studies:

\begin{itemize}
  \item Surface adsorption under an applied bias, demonstrating how the adsorption energy can be tuned by the external field.
  \item Field‑ion microscopy simulations, where the impact of external fields on atom diffusion and desorption is captured accurately.
  \item Electrochemical interfaces, showing the ability to model charged electrodes and to explore potential‑dependent reaction pathways.
  \item Implicit‑solvation and QM/MM setups, in which the solvent is represented by an effective potential \(V^{\text{solvent}}\) (i.e., the averaged electrostatic potential generated by the MM charge density) that is added to \(V_{\text{ext}}\).
\end{itemize}

By exposing the external potential as a user‑controlled input, the implementation provides full control over the electrostatic environment while retaining the standard workflow of VASP calculations.  Consequently, it offers a versatile and easily extensible framework for investigating a wide range of field‑induced phenomena — from catalysis and corrosion to nanoscale device physics —within the well‑established DFT paradigm.

\section{Acknowledgments}
We acknowledge funding by the Deutsche Forschungsgemeinschaft (DFG, German Research Foundation) through SFB1394, project no. 409476157 and SFB1625, project no. 506711657. J. Y. acknowledges support by the Alexander von Humboldt foundation. 
S. M. acknowledges financial support from the International Max Planck Research School for Sustainable Metallurgy (IMPRS SusMet).
The authors thank Dr. Sudarshan Vijay for the helpful discussion and technical support of the VASP-Python plugin.  

\section{Data Availability}
The VASP-Python plugin files and examples are available at \url{https://github.com/eisenforschung/VASP-Python}.

\bibliography{references}

\end{document}